\documentclass[twocolumn,prd,amsfonts]{revtex4}
\usepackage[final]{graphicx}
\usepackage{dcolumn}
\usepackage{bm}
\usepackage{longtable}
\usepackage{amsmath}

\newcommand{\be}{\begin{equation}}
\newcommand{\ee}{\end{equation}}
\newcommand{\ba}{\begin{eqnarray}}
\newcommand{\ea}{\end{eqnarray}}
\newcommand{\bs}{\begin{subequations}}
\newcommand{\es}{\end{subequations}}

\renewcommand{\d}{\mathrm{d}}

\begin{document}

\title{The cosmological gravitational wave background from primordial
  density perturbations}

\author{Kishore N. Ananda$^1$, Chris Clarkson$^{1,2}$ and David Wands$^1$}
\affiliation{$^1$ Institute of Cosmology and Gravitation, University of
Portsmouth, Mercantile House, Portsmouth PO1 2EG, United Kingdom\\
$^2$ {Cosmology and Gravity Group,
  Department of Mathematics and Applied Mathematics,
  University of Cape Town, Rondebosch 7701, Cape Town, South Africa}\\
  Email:~\tt{kishore.ananda@port.ac.uk, chris.clarkson@uct.ac.za,
  david.wands@port.ac.uk}}

\date{\today}

\begin{abstract}


We discuss the gravitational wave background generated by primordial
density perturbations evolving during the radiation era. At
second-order in a perturbative expansion, density fluctuations produce
gravitational waves. We calculate the power spectra of gravitational
waves from this mechanism, and show that, in principle, future
gravitational wave detectors could be used to constrain the primordial
power spectrum on scales vastly different from those currently being
probed by large-scale structure.
As examples we compute the gravitational wave background generated by
both a power-law spectrum on all scales, and a delta-function power
spectrum on a single scale.

\end{abstract}

\maketitle

\section{Introduction}

Gravitational waves are an inevitable, yet still elusive,
consequence of Einstein's theory of general relativity that will be
tested and, we hope, revealed by upcoming experiments.
Linear perturbations about a cosmological metric include transverse
and trace-free modes (tensor modes) that propagate independently of
conventional matter perturbations at first order - i.e., gravitational
waves. During an inflationary expansion in the very early universe
large scale (super-Hubble scale) tensor modes will be generated from
initial quantum fluctuations on small scales
\cite{Rubakov:1982df,Fabbri:1983us,Abbott:1984fp}. But the amplitude
depends on the energy scale of inflation and may be unobservably small
if inflation occurs much below the GUT scale \cite{Knox:2002pe}.

On the other hand primordial density perturbations, and their
associated scalar metric perturbations, do exist and these will
inevitably generate a cosmological background of gravitational waves
at second order through mode coupling
\cite{Tomita,Matarrese:1992rp,Matarrese:1993zf,Matarrese:1997ay,Noh:2004bc,Carbone:2004iv}. Given
the detailed information we now have about the primordial density
perturbations on a range of cosmological scales, it is now timely to
consider the amplitude and distribution of tensor (and vector) modes
that will be generated at second order.

The observed Gaussian distribution of the primordial density
perturbations will generate second-order modes, which will have a
$\chi^2$-distribution, unlike any first-order gravitational waves from
inflation. Given the observed primordial power spectrum of scalar
perturbations on large scales, of order $10^{-9}$ \cite{WMAP3}, one
would expect the power spectrum of second-order metric perturbations to be
of order $10^{-18}$ in the radiation dominated era. But this naive
expectation needs to be tested against a full second-order
calculation. To our knowledge this has not previously been done.
Recently the effect on the cosmic microwave background (CMB) of
second-order gravitational waves generated on very large scales was
investigated by Mollerach {\em et al} \cite{Mollerach:2003nq}
but these modes enter the Hubble scale after matter-radiation
equality.  We will consider much smaller scale modes that
enter the Hubble scale during the primordial radiation-dominated
era and may be relevant for direct detection by gravitational wave
experiments.

Moreover, the generation of gravitational waves from primordial
density perturbations on smaller scales, not directly probed by
astronomical observations, could be used in the same way that
primordial black hole formation has previously been used to constrain
overdensities on these scales \cite{Liddle:1998nt}.  Recently Easther
and Lim \cite{Easther:2006gt} (see also
Ref.\cite{Khlebnikov,Bassett:1997ke,Felder:2006cc}) have suggested
that large density inhomogeneities on sub-Hubble scales during
preheating at the end of inflation could generate a gravitational wave
background that might be detectable by future gravitational wave
experiments such as Advanced LIGO \cite{ALIGO}.

In this paper we present the second-order evolution equation for
gravitational waves generated from terms quadratic in the
first-order matter and metric perturbations.
%
%
In section \ref{background} we present the field equations for the
background Friedmann-Lema\^\i tre-Robertson-Walker (FLRW) metric and
first-order perturbations, giving the standard solutions in terms of
the Bardeen metric potential during the primordial radiation-dominated
era. In section \ref{secondorder} we present the second-order
evolution equation for gravitational waves driven by a source which is
quadratic in the first-order scalar perturbations. We use the
first-order constraint equations to eliminate the matter perturbations
and write this evolution equation solely in terms of the Bardeen
potentials and their derivatives. Finally we present solutions to the
second-order gravitational wave equation using a Green function
method. We calculate the power spectrum of the gravitational waves
generated first by a delta-function spectrum of density perturbations
at a particular wavelength, and then by a power-law spectrum for the
primordial density perturbations. We conclude by discussing the expected
amplitudes of the gravitational waves produced and compare with
sensitivities of current and future detectors in section
\ref{discuss}.

\section{density fluctuations in the radiation era}\label{background}

We will consider perturbations up to second order about a
Robertson-Walker background.  We decompose the metric as
\footnote{where Greek indices run from $0\ldots3$ and Latin indices
run from $1\ldots3$.}
\begin{eqnarray}
\bar{g}_{\alpha\beta} &=& g_{\alpha\beta} + \delta g_{\alpha\beta}
+ \delta^2 g_{\alpha\beta}.
\end{eqnarray}
As we are considering gravitational waves sourced by first-order
scalar perturbations, $\delta g_{\alpha\beta}$ has purely scalar
degrees of freedom, while $\delta^2 g_{\alpha\beta}$ has, in
general, scalar, vector and tensor modes induced by $\delta
g_{\alpha\beta}$. However, we are only investigating second-order
tensor modes so will project out any scalar or vector modes at
second order; we may therefore consider $\delta^2 g_{\alpha\beta}$
as having purely tensor degrees of freedom. Choosing a
longitudinal gauge at first order, we write the metric as
\begin{eqnarray}
\bar{g}_{00} &=& -a^2\left[ 1 + 2\Phi\right],~~~~~
\bar{g}_{0i} = 0,\nonumber\\
\bar{g}_{ij} &=& a^2\left[ \left(1-2\Phi\right)
\gamma_{ij} +\frac{1}{2}h_{ij} \right].
\end{eqnarray}
Here, $\Phi$ is the first-order Bardeen potential, and $h_{ij}$ is
the second-order tensor mode. Note that $\gamma^{jk}h_{ij|k}=0$
and $\gamma^{ij}h_{ij}=0$.

We assume a spatially flat geometry and a pure radiation
background. The scale factor, Hubble rate and energy density
evolve as $a = a_{0}\left(\frac{\eta}{\eta_{0}}\right)$, ${\cal H}
= aH= {\eta}^{-1}$ and, $\rho\propto \eta^{-4}$, in terms of
conformal time $\eta$.

At first-order, assuming no anisotropic pressure, the Bardeen
potential, for a comoving wavenumber $k$, satisfies
\begin{eqnarray}
\Phi''+3{\cal H}\left(1+c_{s}^{2}\right)\Phi'+c_{s}^{2}k^{2}
\Phi &=& 0,
\end{eqnarray}
where the speed of sound is $c_s^2=1/3$, and a prime denotes a
derivative with respect to conformal time. The general solution to
this is
\begin{eqnarray}\label{Phi}
\Phi({\bm k},\eta)&=& \frac{{A}({\bm k})}{(k\eta)^3} \left[
\frac{k\eta}{\sqrt{3}}\cos
\left(\frac{k\eta}{\sqrt{3}}\right)-\sin
\left(\frac{k\eta}{\sqrt{3}}\right) \right] \nonumber\\&+&
\frac{{B}({\bm k})}{(k\eta)^3} \left[ \frac{k\eta}{\sqrt{3}}\sin
\left(\frac{k\eta}{\sqrt{3}}\right)+\cos
\left(\frac{k\eta}{\sqrt{3}}\right) \right].
\end{eqnarray}
This will act as a source for the GW's at second order. At early
times, $k\eta\to0$ we see that
\begin{equation}
\Phi({\bm
k},\eta)=-\frac{A(\bm k)}{9\sqrt{3}}+\frac{B(\bm k)}{(k\eta)^3}
;
\end{equation}
the second term is the decaying mode which we shall neglect hereafter.

Assuming that the fluctuations are Gaussian, we may write
${A}({\bm k}) = {A}(k) \hat{E}({\bm k})$ where the $\hat{E}$ are
Gaussian random variables of unit variance which have the property
\begin{eqnarray}
\langle \hat{E}^{*}({\bm k}_{1})
\hat{E}({\bm k}_{2}) \rangle &=&
\delta^{3}({\bm k}_{1}-{\bm k}_{2})  .
\end{eqnarray}
The power spectrum for the scalar perturbation can then be defined
as
\begin{eqnarray}
\langle \Phi^{*}({\bm k}_{1})
\Phi({\bm k}_{2}) \rangle &=& \frac{2\pi^2}{k^3}
\delta({\bm k}_{1}-{\bm k}_{2})~\mathcal{P}_{\Phi}(k,\eta),
\end{eqnarray}
implying
that at early times the power spectrum becomes
\begin{eqnarray}
\mathcal{P}_{\Phi}(k) &\simeq& {A}(k)^2
\frac{k^3}{486\pi^2} \,.
\end{eqnarray}
The Bardeen potential can be related to the comoving curvature
perturbation at early times, giving
\begin{eqnarray}
{A}(k)^2 &\approx& \frac{216\pi^2}{k^3} \Delta_{\mathcal{R}}^{2}(k),
\end{eqnarray}
where $\Delta_{\mathcal{R}}^{2}$ is
primordial power spectrum for the curvature perturbation $\mathcal{R}$.
%
Current observations show $\Delta_{\mathcal{R}}^{2} \simeq
2 \times 10^{-9}$ at a scale $k_{CMB}=0.002 \mathrm{Mpc}^{-1}$,
and is almost independent of wavenumber on these scales~\cite{WMAP3}.

\section{The induced gravitational waves}\label{secondorder}

We now consider the evolution equations for the second order tensor
perturbations, $h_{ij}$, sourced by the scalar density perturbations
discussed above.

We will write the  Fourier transform of $h_{ij}$ as
\begin{eqnarray}
h_{ij}(\bm{x},\eta)& =&  \frac{1}{(2\pi)^{3/2}} \int \d^3
{k}\,e^{i{\bm k}\cdot\bm{x}}\left[ {h}({\bm k},\eta){q}_{ij}({\bm
k}) \right.\nonumber\\&&+\left.
\bar{h}({\bm k},\eta)\bar{q}_{ij}({\bm k}) \right]
,
\end{eqnarray}
where the two polarization tensors $q_{ij}$ and $\bar{q}_{ij}$ are
expressed in terms of the orthonormal basis vectors
${\bm e}$ and $\bar{\bm e}$
orthogonal to $\bm k$,
\begin{eqnarray}
q_{ij}({\bm k}) &=& \frac{1}{\sqrt{2}}\left[ {e}_{i}({\bm
k}){e}_{j}({\bm k}) -\bar{e}_{i}({\bm k})\bar{e}_{j}({\bm
k})\right],
\nonumber\\
\bar{q}_{ij}({\bm k})&=& \frac{1}{\sqrt{2}}\left[
{e}_{i}({\bm k})\bar{e}_{j}({\bm k}) +\bar{e}_{i}({\bm
k}){e}_{j}({\bm k})\right].
\end{eqnarray}

\begin{widetext}

\noindent
Thus to extract the transverse, trace-free part of any tensor we
project with the operator $\hat{\mathcal{T}}_{ij}{}^{lm}$, defined through its action on a two-index tensor
\begin{eqnarray}\label{tenProjTensor}
\hat{\mathcal{T}}_{ij}{}^{lm} {\mathcal S}_{lm} &=&\int d^3 {k}'~ \frac{q_{ij}({\bm
k}')}{(2\pi)^{3/2}} ~\int d^3 {x}'~
\frac{q^{lm}({\bm k}')}{(2\pi)^{3/2}}e^{i{\bm k}'\cdot({\bm x}-{\bm x}')}
{\mathcal S}_{lm}({\bm x}')
+\int d^3 {k}'~ \frac{\bar{q}_{ij}({\bm k}')}{(2\pi)^{3/2}} ~\int d^3 {x}'~ \frac{\bar{q}^{lm}({\bm k}')}{(2\pi)^{3/2}}
e^{i{\bm k}'\cdot ({\bm x}-{\bm x}')} {\mathcal S}_{lm}({\bm x}')
.\nonumber\\
\end{eqnarray}

The evolution equation is calculated by expanding the Einstein field
equation up to second order including quadratic terms in the
first-order scalar modes. The only contributions from pure second
order matter fluctuations would come from a transverse and traceless
contribution to the anisotropic stress which we ignore in this
analysis. Thus, calculating the transverse, trace-free spatial part of
the field equations yields~\cite{thesis,ACBW}
\begin{eqnarray}
 h_{ij}'' + 2{\cal H} h_{ij}'
-\nabla^{2} h_{ij}  = -4
\hat{\mathcal{T}}_{ij}{}^{lm}\mathcal{S}_{lm} .
\end{eqnarray}

The source term is given by~\cite{thesis,ACBW}
\begin{eqnarray}
\mathcal{S}_{ij} &=&
 4\Phi\Phi_{|ij}
+2\Phi_{|i}\Phi_{|j} -
\frac{3}{\kappa^{2}a^2 \rho}
\left[ {\cal H}^{2}\Phi_{|i}\Phi_{|j} +2\mathcal{H}\Phi_{|i}\Phi_{|j}' +
\Phi_{|i}'\Phi_{|j}' \right].
\end{eqnarray}
A pipe denotes the spatial covariant derivative, and
$\kappa^2=8\pi G$.
This expression is consistent with the second-order Ricci tensor for
scalar perturbations about an FRW background calculated in
Ref.~\cite{SOABMR}.


In Fourier space we then find the amplitude of the tensor mode, for either polarisation, obeys the evolution equation
\begin{eqnarray}
h''({\bm k},\eta) + \frac{2}{\eta} h'({\bm k},\eta) +{k}^{2}h({\bm
k},\eta) =
\mathcal{S}({\bm k},\eta),
\end{eqnarray}
where the source term is given by
\begin{eqnarray}
 \label{Sofk}
\mathcal{S}({\bm k},\tilde{\eta}) &=&
\frac{ q^{ij}({\bm k})}{(2\pi)^{3/2}} \int d^3
\tilde{{k}}~
\tilde{{k}}_{i}\tilde{{k}}_{j}~ \left\{
12\Phi({\bm k}-\tilde{{\bm k}},\tilde{\eta}) \Phi(\tilde{{\bm
k}},\tilde{\eta}) +\left[
\tilde{\eta}\Phi({\bm k}-\tilde{{\bm k}},\tilde{\eta})
+\frac{\tilde{\eta}^2}{2}\Phi'({\bm k}-\tilde{{\bm k}},
\tilde{\eta})\right] \Phi'(\tilde{{\bm k}},\tilde{\eta})
\right\}.
\end{eqnarray}
The particular solution for the gravitational waves is then given
by an integral over the Greens function
\begin{eqnarray}
{h}({\bm k},\eta)&=& \frac{1}{a(\eta)}\int_{\eta_{0}}^{\eta}
G_{k}(\eta,\tilde{\eta}) a(\tilde\eta) \mathcal{S}({\bm
  k},\tilde{\eta})d\tilde{\eta},
\end{eqnarray}
where
\begin{eqnarray}
G_{k}(\eta,\tilde{\eta})&=& \frac{4}{\pi^2k} \left[\frac{}{}
\sin(k\eta)\cos(k\tilde{\eta})  -\cos(k\eta)
\sin(k\tilde{\eta}) \right].
\end{eqnarray}

The power spectrum of the induced GW is defined in the usual
manner,
\begin{eqnarray}
\langle {h}({\bm k}_{1},\eta)
{h}({\bm k}_{2},\eta) \rangle = \frac{2\pi^2}{k^3}
\delta({\bm k}_{1}-{\bm k}_{2})~
\mathcal{P}_{{h}}(k,\eta) .
\end{eqnarray}
Substituting for $h({\bm k},\eta)$ and using Wick's theorem, and
using spherical coordinates in Fourier space we find that
\begin{eqnarray}
 \label{integF}
\mathcal{P}_{{h}}(k,\eta) &=& \frac{1}{8\pi^4 a(\eta)^2}
\int^{\eta}_{\eta_{0}} d\tilde{\eta}_{2} \int^{\eta}_{\eta_{0}}
d\tilde{\eta}_{1}~a(\tilde{\eta}_{1}) a(\tilde{\eta}_{2})~
\int^{\infty}_{0} d\tilde{k}
\int^{u_+}_{u_-}
du~k^{2}G_{k}(\eta,\tilde{\eta}_{1}) G_{k}(\eta,\tilde{\eta}_{2})
F(\tilde{k},k,u;\tilde\eta_1,\tilde\eta_2).
\end{eqnarray}
The variable $u$ is given by
\begin{eqnarray}\label{var1}
u = \sqrt{1+(\tilde{k}/k)^2-2(\tilde{k}/k)\cos\theta},
\end{eqnarray}
where $\theta$ is the angle between the modes $\bm k$ and
$\tilde{\bm k}$,  and the limits of integration $u_\pm$ correspond
to the angles $\theta=0,\pi$ respectively.  The integrand $F$ in
Eq.~(\ref{integF}) is found to be
\begin{eqnarray}
 \label{integrandF}
F(\tilde{k},k,u;\tilde\eta_1,\tilde\eta_2) &=& u\tilde{k} \left[
(2k\tilde{k})^2-((uk)^2-k^2-\tilde{k}^2)^2 \right]^2\nonumber\\
&\times& \left\{
3\Phi(uk,\tilde{\eta}_1)\Phi(\tilde{k},\tilde{\eta}_1)
+\left[ 2\tilde{\eta}_1\Phi(uk,\tilde{\eta}_1) +
\tilde{\eta}_{1}^{2}\Phi'(uk,\tilde{\eta}_1) \right]
\Phi'{}(\tilde{k},\tilde{\eta}_1) \right\} \nonumber\\
 &\times & \left\{
3\Phi(uk,\tilde{\eta}_2)\Phi(\tilde{k},\tilde{\eta}_2)
+\tilde{\eta}_{2}^{2}\Phi'(uk,\tilde{\eta}_2)
\Phi'(\tilde{k},\tilde{\eta}_2) +\tilde{\eta}_2 \left[
\Phi(uk,\tilde{\eta}_2)\Phi'(\tilde{k},\tilde{\eta}_2) +
\Phi(\tilde{k},\tilde{\eta}_2)\Phi'(uk,\tilde{\eta}_2) \right]
\right\}.
\end{eqnarray}
where we distinguish between the Gaussian random variable $\Phi({\bm
k},\eta)$ in Eq.~(\ref{Sofk}) and its amplitude
$\Phi(k,\eta)$ in Eq.~(\ref{integrandF})
which we take to be isotropic. We introduce the following
dimensionless variables
\begin{eqnarray}
 \label{var2}
 v = \frac{\tilde{k}}{k}, \qquad x= k\eta
\end{eqnarray}
and substitute for the first-order solution for $\Phi$ from
Eq.~(\ref{Phi}), as well as the Greens function, into our GW power
spectrum. After some simplification and substitution from the
above formulas we have
\begin{eqnarray}\label{keyint}
\mathcal{P}_{{h}}(k,\eta) &=& \frac{2(216)^2}{\pi^4 \eta^2}
\int^{\infty}_{0} d v~ \int^{|v+1|}_{|v-1|} du~
\frac{1}{(uv)^8}\left[ 4v^2-(u^2-v^2-1)^2 \right]^2
\mathcal{P}_{\Phi}(u k) \mathcal{P}_{\Phi}(v k)\nonumber\\
&& \times \left[ \sin(x) \int^{x}_{x_{0}} d\tilde{x}_{1}
~\mathcal{I}_{1}(\tilde{x}_{1}) -\cos(x) \int^{x}_{x_{0}}
d\tilde{x}_{1} ~\mathcal{I}_{2}(\tilde{x}_{1}) \right]
\left[ \sin(x) \int^{x}_{x_{0}} d\tilde{x}_{2}
~\mathcal{I}_{3}(\tilde{x}_{2}) -\cos(x) \int^{x}_{x_{0}}
d\tilde{x}_{2} ~\mathcal{I}_{4}(\tilde{x}_{2}) \right],
\end{eqnarray}
where we have defined the four functions
\begin{eqnarray}\label{intgen}
\mathcal{I}_{j}({x}) &=& \sum_{m=1}^{5}~\sum_{n=1}^{8}~\sin
\left( \alpha_{n}{x} + \phi_{n}\right)
\frac{M_{nm}^{j}}{{x}^{m}}.
\end{eqnarray}
The coefficients $\alpha_n,\phi_n$ and $M_{nm}^j$ in this
expression are dependent on $u$ and $v$ but not $x$, and may be
found in the appendix.

Evaluating the integrals in the rhs of Eq.~(\ref{keyint}) for
various input scalar power spectra will then tell us the power in
each GW mode. This is not particularly simple, so we start by
analytically expanding the integrals over the functions
$\mathcal{I}_j$ by parts up to Si and Ci functions \cite{GR}:
\begin{eqnarray}\label{X}
X_j(u,v,x,x_0)=\int_{x_{0}}^{x}~d\tilde{x}~\mathcal{I}_{j}(\tilde{x})
&=&
\sum_{m=1}^{5}~\sum_{n=1}^{8}~M_{nm}^{j}~\left\{\left[\sum_{k=1}^{m-2}~
\frac{(m-k-2)!}{(m-1)!} \alpha_{n}^{k} \sin \left(
\alpha_{n}\tilde{x} + \phi_{n}+\frac{(k+2)}{2}\pi\right)
\tilde{x}^{(1+k-m)}\right]_{x_{0}}^{x} \right.\nonumber\\
&&- \left.  \frac{\alpha_{n}^{(m-1)}}{(m-1)!}
\int_{x_{0}}^{x}~d\tilde{x}~\frac{1}{\tilde{x}}\sin \left( \alpha_{n}\tilde{x} +
\phi_{n}+\frac{(m+1)}{2}\pi\right)  \right\}.
\end{eqnarray}
The remaining two integrals over Fourier space can now be done
numerically once power spectra for the scalar modes are chosen. We
shall only consider modes which start
their evolution
well outside the Hubble radius, and hereafter set $x_0=0$.

\end{widetext}

\subsection{Gravitational wave generation by a single scalar mode}

Second-order gravitational waves potentially provide a method by which
we could detect a particular scalar mode with excessive power,
compared to the roughly scale-invariant average we observe on large
scales today.
As a precursor to a power-law power spectrum for the scalar modes,
we can investigate how the system reacts to power being put in at
one particular scale. That is, we put power in at a single
wavelength (i.e., a single comoving scale) which is described by an
isotropic distribution of wavevectors (i.e., at all possible
angles). This would be useful when considering preheating for
example which can result in features (large power over narrow range
of scales) in the power spectrum of the input scalar
perturbations~\cite{Khlebnikov,Easther:2006gt}.

We will therefore consider the case of the delta-function power
spectrum, as an idealised limit of a spike in the power spectrum,
or of power being introduced on narrow range of scales above the
roughly scale-invariant primordial spectrum observed. The specific
form we choose is:
\begin{eqnarray}
\mathcal{P}_{\Phi}(k) =\frac{4}{9} \mathcal{A}^2
\Delta_{\mathcal{R}}^2(k_{CMB})~ \delta(k-k_{in}),
\end{eqnarray}
where $\mathcal{A}$ is the amplitude at a single wavenumber,
$k_{in}$, relative to the observed amplitude of the primordial
power spectrum, $\Delta^2_{\mathcal{R}}(k_{CMB})$, at wavenumber
$k_{CMB}\gg k_{in}$.

The power spectrum of the gravitational waves produced by this scalar
mode is then, from Eq.~(\ref{keyint}),
\begin{eqnarray}\label{PowerGen}
\mathcal{P}_{{h}}(k,\eta) &=& \frac{2(216)^2}{\pi^4 \eta^2}
\int^{\infty}_{0} d v' \int^{|v+1|}_{|v-1|} du'
\nonumber\\&&\mathcal{P}_{\Phi}(u' k)~\mathcal{P}_{\Phi}(v'
k)~\mathcal{F}_{\delta}(u',v',x),
\end{eqnarray}
where
\begin{eqnarray}
\mathcal{F}_{\delta}(u,v,x) &=& \frac{1}{(uv)^8}\left[
4v^2-(u^2-v^2-1)^2 \right]^2
\nonumber\\&&\!\!\!\!\!\!\!\!\!\!\!\!\!\!\!\!\!\!\!\!\!\!\!\!\!\!\!\!\!\!\!\!\!\!\!\!\!\!
\times
\left[
\sin(x) X_1 -\cos(x) X_2
\right]\left[ \sin(x) X_3 -\cos(x) X_4 \right].
\end{eqnarray}
We may now evaluate the $u'$ and $v'$ integrals for a delta-function power spectrum, giving
\begin{eqnarray}
\mathcal{P}_{{h}}(k,\eta) &=& \frac{2(216)^2}{\pi^4 \eta^2}
(\mathcal{A}\Delta_{\mathcal{R}})^4 ~\mathcal{F}_{\delta}(u,v,x),
\end{eqnarray}
provided that the following condition holds:
\begin{eqnarray}
u=v=\frac{k_{in}}{k}\geq\frac{1}{2}.
\end{eqnarray}
The inequality arises from the fact that gravitational waves
cannot be excited with more than twice the momentum of the density
perturbations, $k_{in}$.

\begin{figure}[ht]
\includegraphics[width=0.45\textwidth]{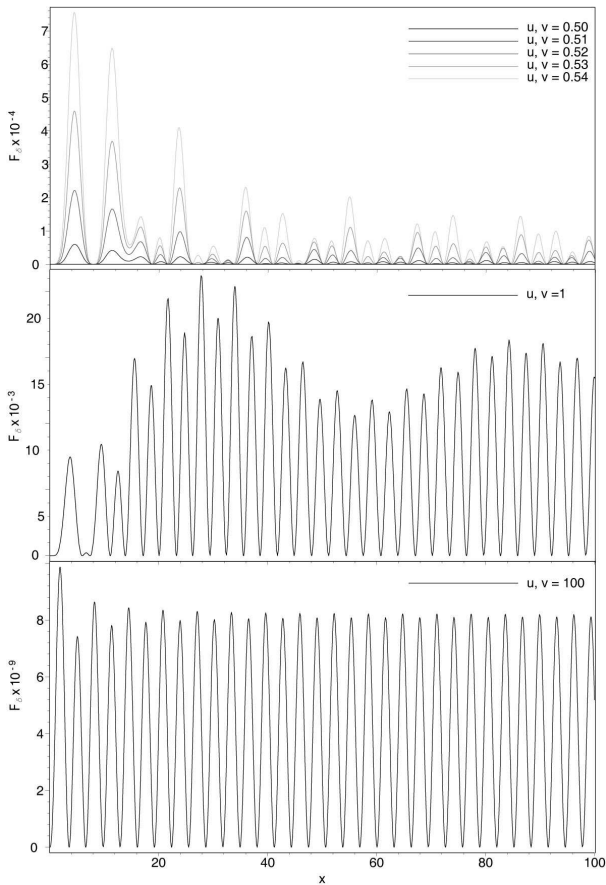}
\caption{The function $\mathcal{F}_{\delta}(x)$ for different values
of $k_{in}/k=u=v$, with $x_0=0$. }\label{fig1}
\end{figure}
\begin{figure}[ht]
\includegraphics[width=0.45\textwidth]{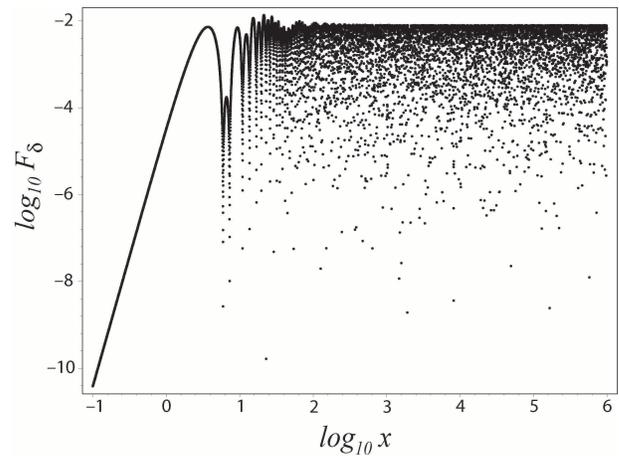}
\caption{The function $\mathcal{F}_{\delta}(x)$ from early times
(super-Hubble scales) to late times (small scales) for $k=k_{in}$
(i.e. $u=v=1$), $x_0=0$. For large $x$ the function oscillates with
a constant amplitude.}\label{fig2}
\end{figure}
\begin{figure}[ht]
\includegraphics[width=0.45\textwidth]{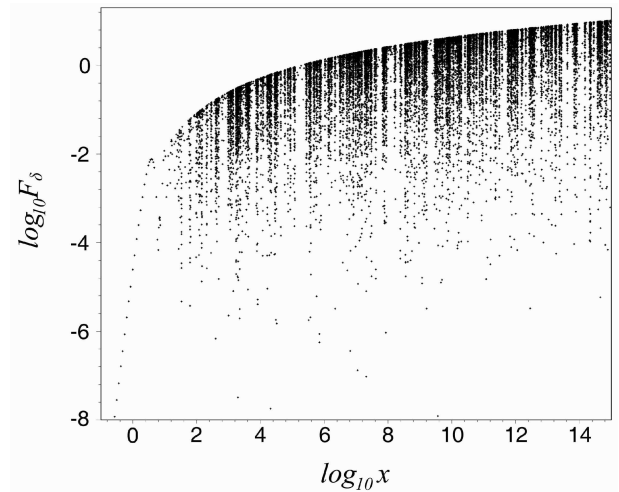}
\caption{The function $\mathcal{F}_{\delta}(x)$ from early times
(super-Hubble scales) to late times (small scales), for the resonant
case of $k_{in}/k=u=v=\sqrt{3}/2$. For large $x$ the power in the
generated GW continues to grow logarithmically. }\label{fig2b}
\end{figure}
\begin{figure}[ht]
\includegraphics[width=0.45\textwidth]{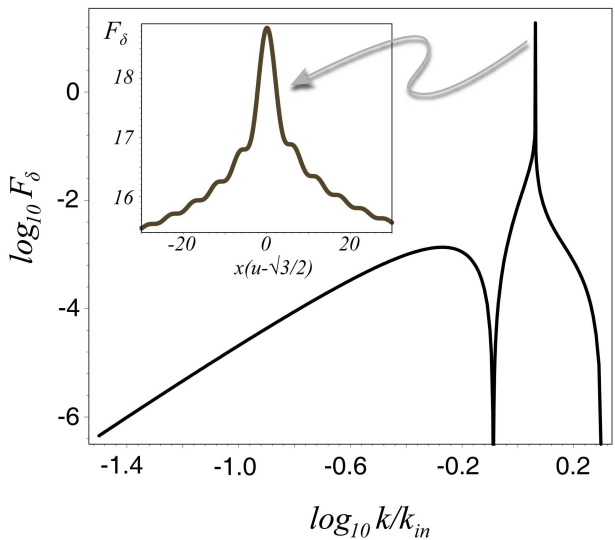}
\caption{ The power at late times, $x=10^{20}$, for all scales,
showing a resonance at $u=\sqrt{3}/2$ and a power-law tail (due to
modes on larger scales being uncorrelated). Detail of the resonance
is shown in the inset revealing a wobbly structure.}\label{fig3}
\end{figure}

Although $\mathcal{F}_{\delta}(v,v,x)$ is a very complicated
function to write down, its properties are relatively
straightforward to understand, as we can see from
Figs.~\ref{fig1}~-- \ref{fig3}.

In Fig.~\ref{fig1} we see the basic behaviour for small $x$. When
$x_0=0$, we may interpret $x$ as time for a fixed wavelength or
wavenumber at fixed time. The function therefore is zero at early
times and large scales, has interesting oscillatory behaviour on
scales of order the horizon size, when $k\eta=1$, and oscillates at
constant amplitude at small scales and/or late times~-- i.e., when
modes are well inside the horizon. The asymptotic behaviour is
clearly seen in Fig.~\ref{fig2}, where we are looking at modes with
$k=k_{in}$, generated when there is a 60 degree angle
($\theta=60^{\circ}$) between the input and output mode (where
$\theta$ is given by Eq.~\ref{var1} and Eq.~\ref{var2}). We have
power-law growth for small $x$, and constant amplitude for
$x\to\infty$, with interesting oscillations on scales of the order
of the Hubble scale. In Fig.~\ref{fig2b} we show the same but now
with $u=v=\sqrt{3}/2$. We see that the amplitude continues to grow
on small scales or late times, corresponding to resonant
amplification of modes generated at an angle $\cos\theta=1/\sqrt{3}$
to the incoming mode.

In Fig.~\ref{fig3} we show the envelope of $\mathcal{F}_{\delta}$ at
fixed large $x$ (where we have chosen $x=10^6$) which shows the
shape of the power spectrum one could observe at late times. There
is a power-law tail on large scales, $k \ll k_{in}$ which is
proportional to $(k/k_{in})^3$ as one would expect for perturbations generated
from much smaller scale modes and hence uncorrelated on larger scales. On
small scales there is a sharp high-frequency cutoff at $k=2k_{in}$.
We also see the resonance as a sharp spike in the power spectrum, of
approximate width $1/x$ and amplitude
$0.05(\log_{10}x)(\log_{10}0.02x)$.
Finally we note there is zero power at $u=v=\sqrt{3/2}$, corresponding
to an angle $\cos\theta=1/\sqrt{6}$ between the input and generated
modes.

\subsection{Power law scalar modes}

\begin{figure}[t]
\includegraphics[width=0.45\textwidth]{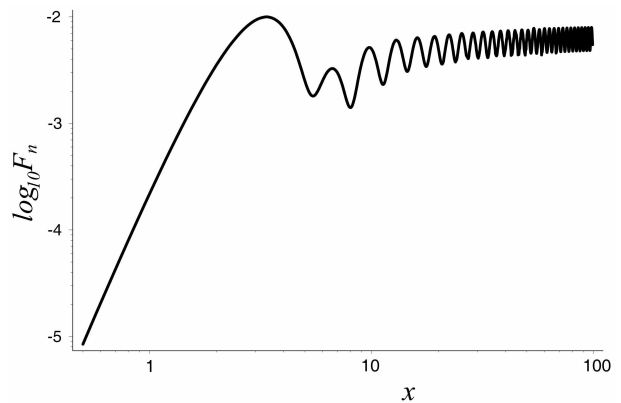}
\caption{ $\mathcal{F}_{n_{s}}(x)$ for a scale invariant input power
spectrum. For this case the power spectrum of gravitational waves is
also scale invariant owing to the scale invariant conversion  factor
between $\mathcal{F}_{n_{s}}$ and $\mathcal{P}_{h}$ given by
Eq.~(\ref{ph-fn}).}\label{fig4}
\end{figure}

While the preceding subsection gave insight into many aspects of the
generation mechanism, we would also like to know the GW generated from
a nearly scale-invariant spectrum of density fluctuations. To
investigate this, we assume that the input power spectrum is:
\begin{eqnarray}
\mathcal{P}_{\Phi}(k_{in}) =\frac{4}{9}
\Delta_{\mathcal{R}}^2~\left(\frac{
k_{in}}{k_{CMB}}\right)^{n_{s}-1},
\end{eqnarray}
where the index $n_s$ tells us the tilt of the spectrum relative to
scale-invariance, $n_s=1$, and $k_{CMB}$
is a pivot scale for the power spectrum~\cite{WMAP3}. The power spectrum
of the generated gravitational waves is then given, from
Eq.~\ref{PowerGen}, by \be\label{ph-fn} \mathcal{P}_{h}(k,\eta) =
\frac{2(216)^2\Delta_{\mathcal{R}}^4}{\pi^4
\eta^2}~\left(\frac{k}{k_{CMB}}\right)^{2(n_{s}-1)}~\mathcal{F}_{n_{s}}(x).
\ee The function $\mathcal{F}_{n_{s}}$ is defined by
\be
\!\!\!\mathcal{F}_{n_{s}}(x) = \int^{\infty}_{0} d v~
\int^{|v+1|}_{|v-1|} du~(u v)^{n_{s}-1}~\mathcal{F}_{\delta}(u,v,x).
\ee
We show a plot of this function for a scale-invariant spectrum,
$n_s=1$ in Fig.~\ref{fig4}.

We can think of $\mathcal{F}_{n_{s}}(x)$ as a function of wavenumber
$k=x/\eta$ at a specific time $\eta$, in which case the amplitude of
$\mathcal{F}_{n_{s}}$ peaks on scales just inside the Hubble radius,
and becomes scale-invariant on smaller scales. The series of
oscillations for $x>1$ is analogous to the acoustic peaks in the CMB
spectrum, although one might have expected this to be less pronounced
at second order where there is an integration over modes.  It suggests
that a narrow range of 1st-order modes with well-defined phase make
the dominant contribution to the second-order tensors on a given scale.

On the other hand, Fig.~\ref{fig4} also represents the evolution of a
single mode over time. Thus, power is continually added to the GW
until it is well inside the Hubble radius, before oscillating at
almost constant amplitude at late times, i.e., a freely propagating GW.

\begin{figure}[ht]
\includegraphics[width=0.45\textwidth]{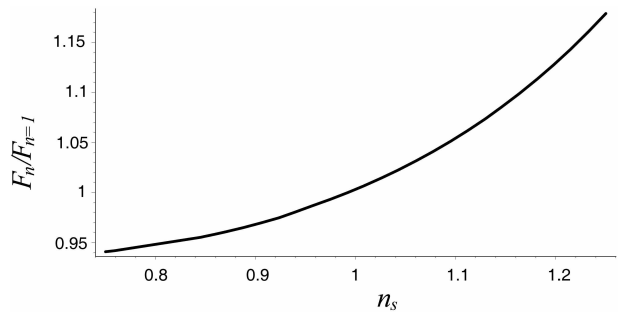}
\caption{ $\mathcal{F}_{n_{s}}$ for  $x\to\infty$, $x_0=0$, vs. the tilt $n_s$ of the scalar power spectrum. We have normalised the vertical scale to
$\mathcal{F}_{n_{s}=1}=8.3\times10^{-3}$. }\label{fig5}
\end{figure}

How does the tilt of the scalar modes affect the generated GW? As
far as $\mathcal{F}_{n_{s}}$ is concerned there is only a small
amplification change, shown in Fig.~\ref{fig5} for large $x$.
There is also an increase in the power at small scales due to the
factor $(k/k_{CMB})^{2(n_{s}-1)}$ in Eq.~\ref{ph-fn} if we consider
a blue spectrum ($n_{s}>1$).

Finally, we note that for non-zero $x_0$ there will be a downward
break for large $k$ in the scale invariant $F_{n_s}$, which
corresponds to $k\eta_0\sim1$~-- i.e., modes which are inside the
horizon at the start of the interaction period. However, if we
start our integration at the GUT scale this will only be relevant
for very high frequencies ($\gtrsim10^8$Hz), so we shall not
pursue this further here.

\section{Discussion}\label{discuss}


We have calculated the stochastic background of gravitational waves
predicted at second-order due to primordial density perturbations and
scalar metric perturbations at first order in the early
(radiation-dominated) universe. In particular we have
evaluated the power spectrum of gravitational waves produced first, by
a delta-function power spectrum, representing an infinitely sharp
feature in the primordial spectrum at a single wavelength, and
secondly by a power-law spectrum, compatible with direct observations
of the primordial density perturbations on much larger (CMB) scales.
It is interesting to compare these predictions with the sensitivities
of current or planned gravitational wave experiments.

The resulting power spectrum of gravitational waves at
matter-radiation equality, $\eta_{eq}$, can be related to their energy
density (per logarithmic interval) today by~\cite{maggiore}
\ba
 \label{omgw}
\Omega_{GW}(k,\eta)&=&\frac{\Omega_{\gamma}(\eta)}{\Omega_{\gamma}
(\eta_{eq})}
\frac{4\pi}{3}\eta_{eq}^2\mathcal{P}_{{h}}(k,\eta_{eq})\nonumber\\
\ea
using the fact that the energy densities of GW and radiation
evolve at the same rate in the matter era. We shall consider first the
case of the power-law spectrum of primordial density perturbations,
for which we obtain
\ba
\Omega_{GW}(k,\eta)&=&\frac{\Omega_{\gamma}(\eta)}{\Omega_{\gamma}(\eta_{eq})}
\frac{124416}{\pi^3}{\Delta_\mathcal{R}^4}\nonumber\\
&\times&
\left(\frac{k}{k_{CMB}}\right)^{2(n_{s}-1)}~\mathcal{F}_{n_{s}}(k\eta_{eq})
\ea
We have assumed that all modes are outside the Hubble radius at
the start of their evolution, $x_0\ll1$.

The latest WMAP data gives
$\Delta^2_\mathcal{R}(k_{CMB}=0.002\text{Mpc}^{-1})\approx
2.36\times10^{-9}$. Combining this with
$\Omega_{\gamma}\approx5\times10^{-5}$ today, we have
\be
\Omega_{GW}(k,\eta)
\approx2.23\times10^{-18}
\left(\frac{k}{k_{CMB}}\right)^{2(n_{s}-1)}~\mathcal{F}_{n_{s}}(k\eta_{eq})
\ee
The mode which enters the Hubble radius at matter radiation equality
defines $\eta_{eq}=1/k_{eq}$ where $k_{eq}=0.009$Mpc$^{-1}$,
corresponding to a GW frequency today of $f_{eq}=c
k_{eq}/2\pi\approx1.4\times10^{-17}$Hz. Our pivot value corresponds
to frequency $f_{CMB}\approx3.1\times10^{-18}$Hz. For the power-law
case we then have \be \Omega_{GW}\approx
1.86\times10^{-20+34(n_s-1)}\left(3.2\frac{f}{\text{Hz}}\right)^{2(n_s-1)}
\left(\frac{\mathcal{F}_{n_{s}}}{\mathcal{F}_{n_{s}=1}}\right) \ee
where $\mathcal{F}_{n_{s}}$ is evaluated at $x=\infty$, and may be
estimated from Fig.~\ref{fig5}. Therefore, for a red spectrum with
$n_s=0.95$, we find
$\Omega_{GW}\approx3.2\times10^{-22}\left({f}/{\text{Hz}}\right)^{-0.1}$,
while for a blue spectrum with $n_s=1.1$ we have
$\Omega_{GW}\approx6.1\times10^{-17}\left({f}/{\text{Hz}}\right)^{0.2}$.
In principle, therefore, proposed detectors such as LISA~\cite{LISA} or
DECIGO~\cite{BBOseto,UDECkudoh} could be used to place limits on
the scalar tilt, independently of other observations. Of course,
we have only considered constant tilt here; running of the
spectral index will change these predictions considerably. Indeed,
as the primordial power spectrum is directly constrained only at
very large scales, GWs provide a novel way to probe the primordial
power spectrum at wavelengths some twenty orders of magnitude
smaller than CMB measurements allow.

For the case of excess power in a single mode, we may define
$\Omega_{GW}$ analogously with Eq.~(\ref{omgw}), but near the spike
${\cal P}_h$ is not smooth, so $\Omega_{GW}$ is not necessarily the
actual energy density of GW. However, we will use it to
give a rough indication of power
for comparison against quoted detector sensitivities.
The peak of the spike in the power spectrum occurs
at $f_{peak}=2f_{in}/\sqrt{3}$, which gives
\be
\Omega_{GW}(f_{peak})
\approx 6.0\times10^{-17}\mathcal{A}^4\left(1+0.09\log_{10}
\frac{T_{ent}}{1\text{GeV}}\right) \,.
\ee
We have written this in terms of the temperature at which the
input wavenumber enters the Hubble radius, using~\cite{komatsu}
$f_{in}={c
k_{in}}/{2\pi}\approx10^{-6}\left({T_{ent}}/{1~\text{GeV}}\right)\text{Hz}$
for $f\gtrsim10^{-6}$Hz, which is the range for any detector of
GW.
We note that the typical
resolution of a detector in time $T$ is $\Delta
f\sim1/T$~\cite{maggiore}; so, for a one year observation we have
$\Delta f\sim 10^{-8}$Hz.
For current detectors such as GEO~\cite{GEO}, LIGO~\cite{ALIGO}, TAMA~\cite{TAMA} or VIRGO~\cite{VIRGO} whose optimal frequency is $\sim100$Hz,
excess power at a single wavenumber in the
primordial power spectrum would correspond to a horizon entry
temperature of $T_{ent}\approx10^8$GeV, implying that
$\Omega_{GW}\sim 10^{-16}\mathcal{A}^4$. While this is nominally
seven orders of magnitude below Advanced LIGO's sensitivity,
$\Omega_{GW}\sim 10^{-9}$, it
implies that Advanced LIGO could, in principle, pick up modes with
$\mathcal{A}\sim 100$,
i.e., one hundred times the amplitude observed on large scales. For
detectors such as BBO/DECIGO~\cite{BBOcorbin,BBOphinney}, whose
optimal frequency is $\sim 0.1$Hz,
their improved sensitivity ($\Omega_{GW}\sim 10^{-17}-10^{-15}$)
could constrain excess power with $\mathcal{A}\sim1$ at a single wavenumber
at a Hubble entry temperature $\sim10^5$GeV.

In summary, we have presented the second-order evolution equation for
tensor modes driven by quadratic scalar terms, and have solved this to
obtain the resulting power spectrum of the gravitational waves. By
considering density fluctuations with excess power on a given scale,
as well as a nearly scale-invariant power spectrum, we have shown how
future gravitational wave detectors can place constraints on the
primordial density fluctuations at much
smaller scales
than can be probed by observing the CMB and large-scale structure.

\acknowledgments The authors would like to thank Marco Bruni, Roy
Maartens, Karim Malik and Jean-Philippe Uzan for useful comments and
discussions. KNA acknowledges financial support from PPARC.


\bibliography{Ananda}

\newpage
\appendix

\begin{widetext}
\section{Co-efficient matrices in $\mathcal{P}_h$}

The coefficients which appear in Eq.~(\ref{intgen}) and elsewhere
are presented here. First, we define the column matricies
\begin{equation}
\bm{1}=\left(
\begin{array}{c}
  1 \\
  1 \\
  1 \\
  1
\end{array}\right),~~~
\bm{0}=\left(
\begin{array}{c}
  0 \\
  0 \\
  0 \\
  0
\end{array}\right),~~~
\bm{a}=\left(
\begin{array}{c}
  -1 \\
  -1 \\
  +1 \\
  +1
\end{array}\right),~~~
\bm{b}=\left(
\begin{array}{c}
  -1 \\
  +1 \\
  +1 \\
  -1
\end{array}\right),~~~
\bm{c}=\left(
\begin{array}{c}
  +1 \\
  -1 \\
  +1 \\
  -1
\end{array}\right).
\end{equation}
Then we may write:
\be
\alpha_n=\left(
\begin{array}{c}
 \frac{u}{\sqrt{3}}\bm{1}-\frac{v}{\sqrt{3}}\bm{a}+\bm{c}  \\
 \frac{u}{\sqrt{3}}\bm{1}-\frac{v}{\sqrt{3}}\bm{a}+\bm{c}
\end{array}\right),~~~~~
\phi_n=\frac{\pi}{2}\left(
\begin{array}{c}
 \bm{0}\\
 \bm{1}
\end{array}\right),
\ee
and,
\ba
M_{nm}^{1} &=&\left(
\begin{array}{ccccc}
\bm{0} & \frac{\sqrt{3}}{12}u^2v\bm{1}-\frac{\sqrt{3}}{36}u
v^2\bm{a} & \bm{0} & \frac{\sqrt{3}}{2}(u\bm{a}-v\bm{1}) & \bm{0} \\
\frac{1}{36}u^2v^2\bm{a} &\bm{0}&-\frac{1}{12}(3u^2+v^2)\bm{a}
+\frac{1}{2}uv\bm{1}&\bm{0}&\frac{3}{2}\bm{a}
\end{array}\right),\\
M_{nm}^{2} &=&\left(
\begin{array}{ccccc}
\frac{1}{36}u^2v^2\bm{b} &\bm{0}&-\frac{1}{12}(3u^2+v^2)\bm{b}
+\frac{1}{2}uv\bm{c}&\bm{0}&\frac{3}{2}\bm{b}\\
\bm{0} &
-\frac{\sqrt{3}}{12}u^2v\bm{c}-\frac{\sqrt{3}}{36}u v^2\bm{b} &
\bm{0} & \frac{\sqrt{3}}{2}(-u\bm{b}+v\bm{c}) & \bm{0}
\end{array}\right),\\
M_{nm}^{3} &=&\left(
\begin{array}{ccccc}
\bm{0} & \frac{\sqrt{3}}{12}u^2v\bm{1}-\frac{\sqrt{3}}{36}u
v^2\bm{a} & \bm{0} & \frac{\sqrt{3}}{2}(u\bm{a}-v\bm{1}) & \bm{0} \\
\frac{1}{36}u^2v^2\bm{a} &\bm{0}&
-\frac{1}{6}(u^2+v^2)\bm{a}
+\frac{1}{2}uv\bm{1}&\bm{0}&\frac{3}{2}\bm{a}
\end{array}\right),\\
M_{nm}^{4} &=&\left(
\begin{array}{ccccc}
\frac{1}{36}u^2v^2\bm{b} &\bm{0}&-\frac{1}{6}(u^2+v^2)\bm{b}
+\frac{1}{2}uv\bm{c}&\bm{0}&\frac{3}{2}\bm{b}\\
\bm{0} &
-\frac{\sqrt{3}}{12}u^2v\bm{c}+\frac{\sqrt{3}}{36}u v^2\bm{b} &
\bm{0} & \frac{\sqrt{3}}{2}(-u\bm{b}+v\bm{c}) & \bm{0}
\end{array}\right).
\ea

\end{widetext}

\end{document}